\documentclass[runningheads]{llncs}
\usepackage{graphicx}
\usepackage{multirow}
\usepackage{epsfig}
\usepackage{amsmath}
\usepackage{amssymb}
\usepackage{algorithm}
\usepackage{algpseudocode}
\begin{document}

\title{MI$^2$GAN: Generative Adversarial Network for Medical Image Domain Adaptation using Mutual Information Constraint}

\author{Xinpeng Xie\inst{1}\thanks{This work was done when Xinpeng Xie was an intern at Tencent Jarvis Lab} \and Jiawei Chen\inst{2}\thanks{Equal contribution} \and Yuexiang Li\inst{2} \and Linlin Shen\inst{1} \and Kai Ma\inst{2} \and Yefeng Zheng\inst{2}}
\authorrunning{X. Xie et al.}
\institute{Computer Vision Institute, Shenzhen University, Shenzhen, China \\
\email{llshen@szu.edu.cn} \and
Tencent Jarvis Lab, Shenzhen, China\\
\email{vicyxli@tencent.com}
}

\maketitle

\begin{abstract}
    Domain shift between medical images from multicentres is still an open question for the community, which degrades the generalization performance of deep learning models. Generative adversarial network (GAN), which synthesize plausible images, is one of the potential solutions to address the problem. However, the existing GAN-based approaches are prone to fail at preserving image-objects in image-to-image (I2I) translation, which reduces their practicality on domain adaptation tasks. In this paper, we propose a novel GAN (namely MI$^2$GAN) to maintain image-contents during cross-domain I2I translation. Particularly, we disentangle the content features from domain information for both the source and translated images, and then maximize the mutual information between the disentangled content features to preserve the image-objects. The proposed MI$^2$GAN is evaluated on two tasks---polyp segmentation using colonoscopic images and the segmentation of optic disc and cup in fundus images. The experimental results demonstrate that the proposed MI$^2$GAN can not only generate elegant translated images, but also significantly improve the generalization performance of widely used deep learning networks (e.g., U-Net).
    \keywords{Mutual Information \and Domain Adaptation.}
\end{abstract}

\section{Introduction}
Medical images from multicentres often have different imaging conditions, e.g., color and illumination, which make the models trained on one domain usually fail to generalize well to another. Domain adaptation is one of the effective methods to boost the generalization capability of models. Witnessing the success of generative adversarial networks (GANs) \cite{Goodf01} on image synthesis \cite{Isola01,WangT01}, researchers began trying to apply the GAN-based networks for image-to-image domain adaptation. For example, Chen et al. \cite{ChenC2018} used GAN to transfer the X-ray images from a new dataset to the domain of the training set before testing, which increases the test accuracy of trained models. Zhang et al. \cite{ZhangY2018} proposed a task driven generative adversarial network (TD-GAN) for the cross-domain adaptation of X-ray images. Most of the existing GAN-based I2I domain adaptation methods adopted the cycle-consistency loss \cite{Kim01,Yi01,ZhuJ01} to loose the requirement of paired cross-domain images for training. However, recent studies \cite{HuangS01,Zhang01} proved that cycle-consistency-based frameworks easily suffer from the problem of content distortion during image translation. Let $T$ be a bijective geometric transformation (e.g., translation, rotation, scaling, or even nonrigid transformation) with inverse transformation $T^{-1}$, the following generators $G^{'}_{AB}$ and $G^{'}_{BA}$ are also cycle consistent.
\begin{equation}
    \begin{aligned}
        G^{'}_{AB} = G_{AB}T,\;G^{'}_{BA} = G_{BA}T^{-1}
    \end{aligned}\label{eq_gg}
\end{equation}
where the $G_{AB}$ and $G_{BA}$ are the original cycle-consistent generators establishing two mappings between domains $A$ and $B$. Consequently, due to lack of penalty in content disparity between source and translated images, the content of a translated image by cycle-consistency-based frameworks may be distorted by $T$, which is unacceptable in medical image processing.

To tackle the problem, we propose a novel  {\bf GAN} (MI$^2$GAN) to maintain the contents of {\bf M}edical {\bf I}mage during I2I domain adaptation by maximizing the {\bf M}utual {\bf I}nformation between the source and translated images. Our idea relies on two observations: 1) the content features containing the information of image-objects can be fully disentangled from the domain information; and 2) the mutual information, measuring the information that two variables share, can be used as a metric for image-object preservation. Mutual information constraint has been widely used for various medical image processing tasks, such as image registration \cite{MI2003}. Given two variables $X$ and $Y$, the mutual information $I$ shared by $X$ and $Y$ can be formulated as:
\begin{equation}
    \begin{aligned}
        \mathcal{I}(X;Y) = KL(\mathbb{J}||\mathbb{M})
    \end{aligned}\label{eq_MI}
\end{equation}
where $\mathbb{J}$ and $\mathbb{M}$ are joint distribution and the product of marginals of $X$ and $Y$; $KL$ is the KL-divergence. Specifically, $\mathbb{J} = p(y|x)p(x)$ and $\mathbb{M} = p(y)p(x)$, where $x \in X$ and $y \in Y$; $p(x)$ and $p(y)$ are the distributions of $X$ and $Y$, respectively; $p(y|x)$ is the conditional probability of $y$ given $x$.

Since the posterior probability $p(y|x)$ is difficult to be directly estimate \cite{NIPS2016_6399}, we measure and maximize the MI between source and translated images based on the approach similar to \cite{HjelmICLR2019,Belghazi18}. Specifically, the content features of source and translated images are first extracted by the paired adversarial auto-encoders, which are then fed to a discriminator for the estimation of mutual information. Extensive experiments are conducted to validate the effectiveness of our MI$^2$GAN. The experimental results demonstrate that the proposed MI$^2$GAN can not only produce plausible translated images, but also significantly reduce the performance degradation caused by the domain shift.

\begin{figure}[!tb]
    \begin{center}
        \includegraphics[width=0.7\linewidth]{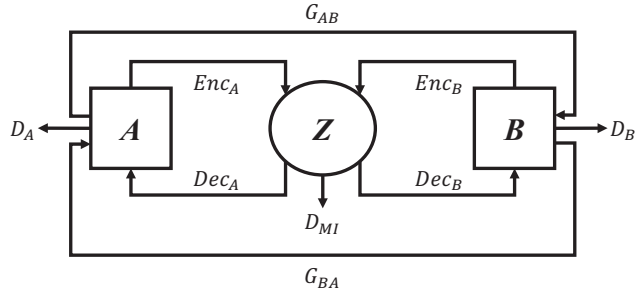}
    \end{center}
    \caption{The framework of our MI$^2$GAN. Similar to CycleGAN \cite{ZhuJ01}, our MI$^2$GAN adopts paired generators ($G_{AB}$ and $G_{BA}$) and discriminators ($D_{B}$ and $D_{A}$) to achieve cross-domain image translation. To preserve image-contents, X-shape dual auto-encoders (\{$Enc_A$, $Dec_A$\} and \{$Enc_B$, $Dec_B$\}) and a mutual information discriminator ($D_{MI}$) are implemented.}
    \label{MutualGAN}
\end{figure}

\section{MI$^2$GAN}
The pipeline of our MI$^2$GAN is presented in Fig.~\ref{MutualGAN}. Similar to current cycle-consistency-based GAN \cite{ZhuJ01}, our MI$^2$GAN adopts paired generators ($G_{AB}$ and $G_{BA}$) and discriminators ($D_{B}$ and $D_{A}$) to achieve cross-domain image translation without paired training samples. To distill the content features from domain information, X-shape dual auto-encoders (i.e., $Enc_A$, $Dec_A$, $Enc_B$, and $Dec_B$) are implemented. The encoders (i.e., $Enc_A$ and  $Enc_B$) are responsible to embed the content information of source and translated images into the same latent space $Z$, while the decoders (i.e.,  $Dec_A$ and  $Dec_B$) aim to transform the embedded content features to their own domains using domain-related information. Therefore, to alleviate the content distortion problem during image translation, we only need to maximize the mutual information between the content features of source and translated images, which is achieved by our mutual information discriminator. In the followings, we present the modules for content feature disentanglement and mutual information maximization in details.

\subsection{X-shape Dual Auto-Encoders}
\begin{figure}[!tb]
    \begin{minipage}[t]{0.49\linewidth}
        \centering
        \includegraphics[width=\linewidth]{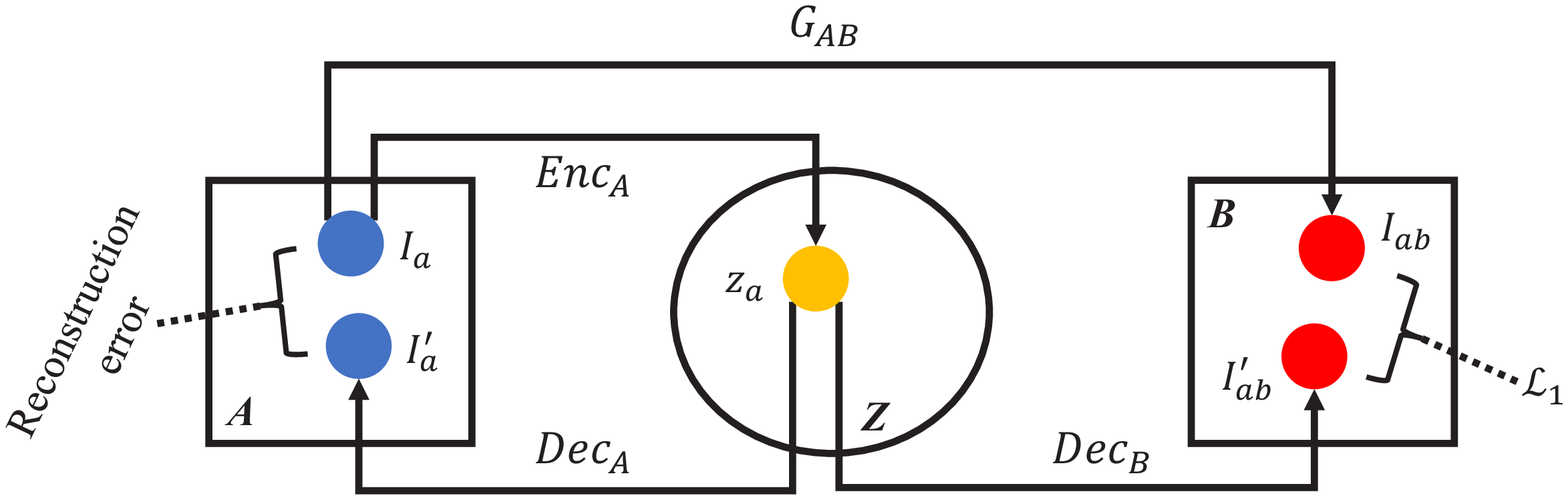}
        {\\(a) Content feature distillation\\}
    \end{minipage}
    \begin{minipage}[t]{0.49\linewidth}
        \centering
        \includegraphics[width=2.0in]{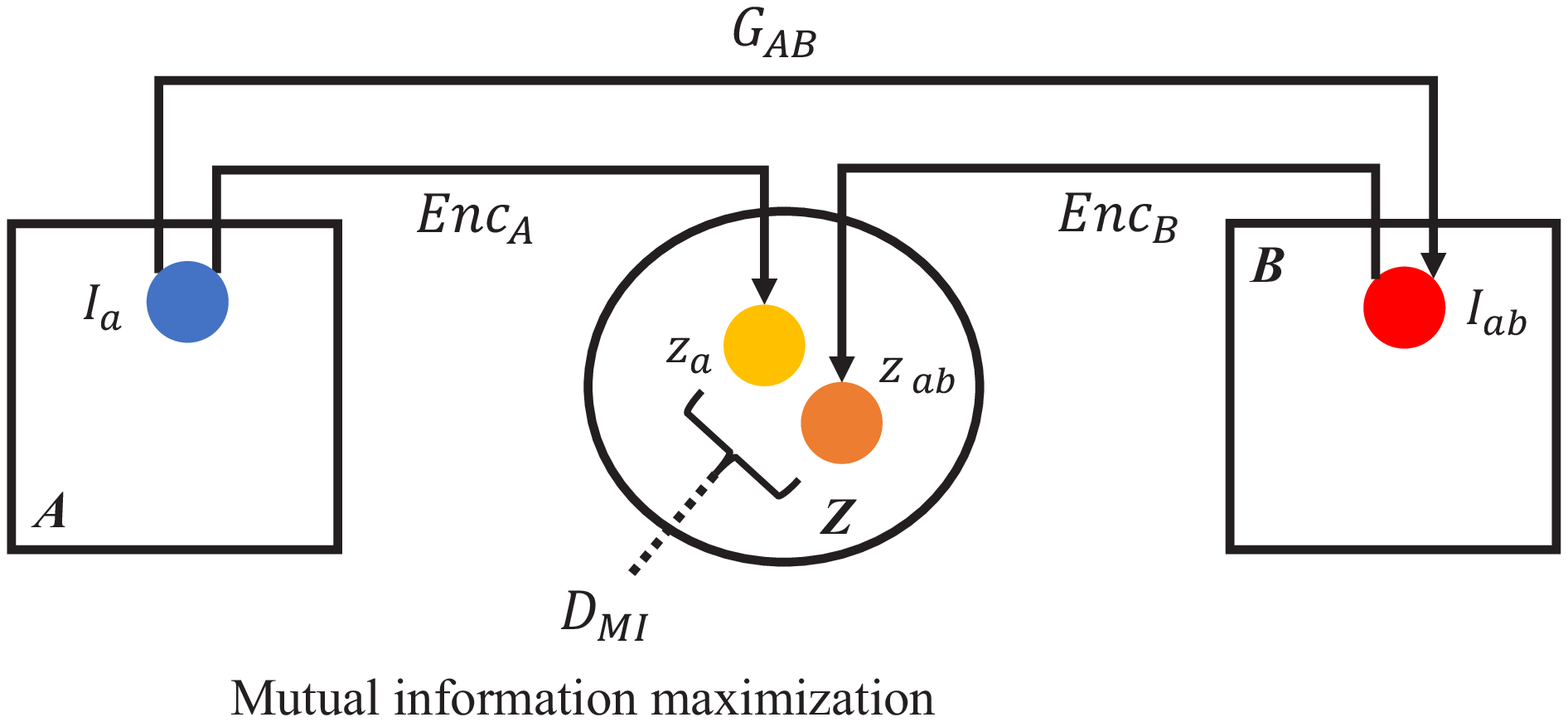}
        {\\(b) Mutual information maximization\\}
    \end{minipage}
    \caption{The pipelines of the main components contained in our framework. (a) X-shape dual auto-encoders and (b) mutual information discriminator.} \label{fig:two_components}
\end{figure}

We proposed the X-shape dual auto-encoders (AEs), consisting of $Enc_A$, $Dec_A$, $Enc_B$, and $Dec_B$, to disentangle the features containing content information. As the mappings between domains A and B are symmetrical, we take the content feature distillation of images from domain A as an example. The pipeline is shown in Fig.~\ref{fig:two_components} (a). Given an input image ($I_a$), the auto-encoder ($Enc_A$ and $Dec_A$) embeds it into a latent space, which can be formulated as:
\begin{equation}
    \begin{aligned}
        z_a = Enc_A(I_a), \;\;\; I_a' = Dec_A(z_a)
    \end{aligned}\label{eq_embedding}
\end{equation}
where $I_{a'}$ is the reconstruction of $I_a$. The embedded feature $z_a$ contains the information of content and domain A. To disentangle them, $z_a$ is mapped to domain B via $Dec_B$:
\begin{equation}
    \begin{aligned}
        I_{ab}' = Dec_B(z_a)
    \end{aligned}\label{eq_embedding_2}
\end{equation}
where $I_{ab}'$ is the mapping result of $z_a$.

As shown in Fig.~\ref{fig:two_components}, apart from the X-shape dual AEs, there is another translation path between domain A and B: $I_{ab} = G_{AB}(I_a)$, where $I_{ab}$ is the translated image yielded by $G_{AB}$. Through simultaneously minimizing the pixel-wise L1 norm between $I_{ab}$ and $I_{ab}'$, and reconstruction error between $I_a$ and $I_{a'}$, $Dec_A$ and $Dec_B$ are encouraged to recover domain-related information from the latent space (in short, the encoders remove domain information and the decoders recover it), which enable them to map the $z_a$ to two different domains. Therefore, the information contained in $z_a$ is highly related to the image-objects without domain bias. The content feature distillation loss ($\mathcal{L}_{dis}$), combining aforementioned two terms, can be formulated as:
\begin{equation}
    \begin{aligned}
        \mathcal{L}_{dis} = ||I_{ab}-I_{ab}'||_1 + ||I_a - I_{a'}||_1.
    \end{aligned}\label{eq_SID}
\end{equation}

\subsection{Mutual Information Discriminator}
Using our X-shape dual AEs, the content features of source $I_a$ and translated $I_{ab}$ images can be disentangled to $z_a$ and $z_{ab}$, respectively. The content feature of translated image preserving image-objects should contain similar information to that of source image. To this end, the encoder ($Enc_B$) needs to implicitly impose statistical constraints onto learned representations, which thereby pushes the translated distribution of $Z_{ab}$ to match the source $Z_{a}$ (i.e., mutual information maximization between $Z_{a}$ and $Z_{ab}$), where $Z_{ab}$ and $Z_{a}$ are two sub-spaces of $Z$.

Referred to adversarial training, which matches the distribution of synthesized images to that of real ones, this can be achieved by training a mutual information discriminator ($D_{MI}$) to distinguish between samples coming from the joint distribution, $\mathbb{J}$, and the product of marginals, $\mathbb{M}$, of the two sub-spaces $Z_a$ and $Z_{ab}$ \cite{HjelmICLR2019}. We use a lower-bound to the mutual information ($\mathcal{I}$ defined in Eq.~\ref{eq_MI}) based on the Donsker-Varadhan representation (DV) of the KL-divergence, which can be formulated as:
\begin{equation}
    \begin{aligned}
        \mathcal{I}(Z_{a} ; Z_{ab}) \geq  \widehat{\mathcal{I}}^{(D V)}(Z_{a} ; Z_{ab})= \mathbb{E}_{\mathbb{J}}\left[D_{MI}(z_{a}, z_{ab})\right]-\log \mathbb{E}_{M}\left[e^{D_{MI}(z_{a}, z_{ab})}\right]
    \end{aligned}
\end{equation}
where $D_{MI}: z_{a} \times z_{ab}\rightarrow \mathbb{R}$ is a discriminator function modeled by a neural network.

To constitute the real ($\mathbb{J}$) and fake ($\mathbb{M}$) samples for the $D_{MI}$, an image is randomly selected from domain B and encoded to $z_{b}$. The $z_{a}$ is then concatenated to $z_{ab}$ and $z_b$, respectively, which forms the samples from the joint distribution ($\mathbb{J}$) and the product of marginals ($\mathbb{M}$) for the mutual information discriminator.

\paragraph{\bf Objective.} With the previously defined feature distillation loss ($\mathcal{L}_{dis}$) and mutual information discriminator, the full objective $\mathcal{L}$ for the proposed MI$^2$GAN is summarized as:
\begin{equation}
    \begin{aligned}
        \mathcal{L} & = \mathcal{L}_{adv}\left(G_{BA},\ D_A\right)+\mathcal{L}_{adv}\left(G_{AB},\ D_B\right) +\alpha \mathcal{L}_{cyc}\left(G_{AB}, \ G_{BA}\right) \\
                    & +\beta \mathcal{L}_{dis}(G_{AB}, Enc_{A}, Dec_{A}, Dec_{B}) +\beta \mathcal{L}_{dis}(G_{BA}, Enc_{B}, Dec_{B}, Dec_{A})                                  \\
                    & + \widehat{\mathcal{I}}(G_{AB}, Enc_{A}, Enc_{B}, D_{MI}) + \widehat{\mathcal{I}}(G_{BA}, Enc_{A}, Enc_{B}, D_{MI})
    \end{aligned}
\end{equation}
where $\mathcal{L}_{adv}$ and $\mathcal{L}_{cyc}$ are adversarial and cycle-consistency losses, the same as that proposed in \cite{ZhuJ01}. The weights $\alpha$ and $\beta$ for $\mathcal{L}_{cyc}$ and $\mathcal{L}_{dis}$ respectively are all set to 10.

\subsection{Implementation Details}
\paragraph{\bf Network architecture.} Consistent to the standard of CycleGAN \cite{ZhuJ01}, the proposed MI$^2$GAN involves paired generators ($G_{AB}$, $G_{BA}$) and discriminators ($D_{B}$, $D_{A}$). Instance normalization \cite{ulyanov2016instance} is employed in the generators to produce elegant translation images, while PatchGAN is adopted in the discriminators \cite{Isola01,li2016precomputed} to provide patch-wise predictions. Our X-shape AEs and mutual information discriminator adopt instance normalization and leaky ReLU in their architectures, and the detailed information can be found in the {\itshape arXiv version}.

\paragraph{\bf Optimization process.} The optimization of $\mathcal{L}_{dis}$ and $\widehat{\mathcal{I}}$ is performed in the same manner of $\mathcal{L}_{adv}$---fixing X-shape dual AEs, $D_{MI}$ and $D_{A}$/$D_{B}$ to optimize $G_{BA}$/$G_{AB}$ first, and then optimize AEs, $D_{MI}$ and $D_{A}$/$D_{B}$ respectively, with fixed $G_{BA}$/$G_{AB}$. Therefore, similar to discriminators, our X-shape dual AEs and mutual information discriminator can directly pass the knowledge of image-objects to the generators, which helps them to improve the quality of translated results in terms of object preservation.

\section{Experiments}
Deep neural networks often suffer from performance degradation when applied to a new test dataset with domain shift (e.g., color and illumination) caused by different imaging instruments. Our MI{$^2$}GAN tries to address the problem by translating the test images to the same domain of the training set. In this section, to validate the effectiveness of the proposed MI{$^2$}GAN, we evaluate it on several publicly available datasets.

\subsection{Datasets}
\paragraph{\bf Colonoscopic datasets.} The publicly available colonoscopic video datasets, i.e., CVC-Clinic \cite{VD01} and ETIS-Larib \cite{SilvaJ01}, are selected for multicentre adaptation. The CVC-Clinic dataset is composed of 29 sequences with a total of 612 images. The ETIS-Larib consists of 196 images, which can be manually separated to 29 sequences as well. Those short videos are extracted from the colonoscopy videos captured by different centres using different endoscopic devices. All the frames of the short videos contain polyps. In this experiment, the extremely small ETIS-Larib dataset (196 frames) is used as the test set, while the relatively larger CVC-Clinic dataset (612 frames) is used for network optimization (80:20 for training and validation).

\paragraph{\bf REFUGE.} The REFUGE challenge dataset \cite{orlando2020refuge} consists of 1,200 fundus images for optic disc (OD) and optic cup (OC) segmentation, which were partitioned to training (400), validation (400) and test (400) sets by the challenge organizer. The images available in this challenge were acquired with two different fundus cameras---Zeiss Visucam 500 for the training set and Canon CR-2 for the validation and test sets, resulting in  visual gap between training and validation/test samples. Since the test set is unavailable, we conduct experiment on I2I adaptation between the training and validation sets. The public training set is separated to training and validation sets according to the ratio of 80:20, and the public validation set is used as the test set.

\paragraph{\bf Baselines overview \& evaluation criterion.} Several unpaired image-to-image domain adaptation frameworks, including CycleGAN \cite{ZhuJ01}, UNIT \cite{LiuMY01} and DRIT \cite{Lee01}, are taken as baselines for the performance evaluation. The direct transfer approach, which directly takes the source domain data for testing without any adaptation, is also involved for comparison. The Dice score (DSC), which measures the spatial overlap index between the segmentation results and ground truth, is adopted as the metric to evaluate the segmentation accuracy.

\subsection{Ablation Study}
\begin{figure}[!tb]
    \begin{center}
        \includegraphics[width=0.85\linewidth]{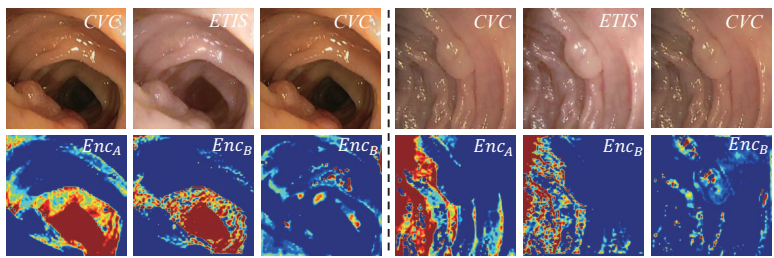}
    \end{center}
    \caption{Content features (the second row) produced by the encoders of our X-shape dual AEs for the input images (the first row) from different domains. The CVC images (left) are manually tuned to ETIS (middle) by experienced experts.}
    \label{fig:content}
\end{figure}

\paragraph{\bf Content feature distillation.} We invite three experienced experts to manually tune two CVC images to the domain of ETIS (as shown in the first row of Fig.~\ref{fig:content}), i.e., tuning the image conditions such as color and saturation based on the statistical histogram of the ETIS domain. The two paired images contain the same content information but totally different domain-related knowledge. To ensure our X-shape dual auto-encoders really learn to disentangle the content features from domain information, we sent the paired images to X-shape dual AEs and visualize the content features produced by $Enc_A$ and $Enc_B$ using CAM \cite{zhou2016learning} (as illustrated in the second row of Fig.~\ref{fig:content}). For comparison, the CVC images are also sent to $Enc_B$ for content feature distillation. It can be observed that the CVC and ETIS images respectively going through $Enc_A$ and $Enc_B$ result in the similar activation patterns, while the encoders yield different patterns for the CVC images. The experimental result demonstrate that the encoders of our X-shape dual AEs are domain-specific, which are able to remove the their own domain-related information from the embedding space.

\paragraph{\bf Mutual information discriminator.} To validate the contribution made by the mutual information discriminator, we evaluate the performance of MI{$^2$}GAN without $D_{MI}$. The evaluation results are presented in Table~\ref{table_colon_simple}. The segmentation accuracy on the test set significantly drops to 65.96\%, 77.27\% and 92.17\% for polyp, OC and OD, respectively, with the removal of $D_{MI}$, which demonstrates the importance of $D_{MI}$ for image-content preserving domain adaptation.

\begin{table}[!tb]
    \centering
    \small
    \caption{DSC (\%) of the polyp segmentation on colonoscopy and the segmentation of optical cup (OC) and optical disk (OD) on REFUGE fundus images, respectively.}\label{table_colon_simple}
    \begin{tabular}{c|c|c|c|c|c|c}
        \hline
        {}                             & \multicolumn{2}{c|}{Colonoscopy} & \multicolumn{4}{c}{Fundus}                                                                               \\\cline{2-7}
        {}                             & {\bf CVC} (val.)                & {\bf ETIS} (test)          & $OC_{val.}$            & $OD_{val.}$            & $OC_{test}$ & $OD_{test}$ \\\hline\hline
        {\bf Direct transfer}          & \multirow{6}{*}{80.79}          & 64.33                      & \multirow{6}{*}{85.83} & \multirow{6}{*}{95.42} & 81.66       & 93.49       \\\cline{1-1}\cline{3-3}\cline{6-7}
        {\bf DRIT} \cite{Lee01}        &                                 & 28.32                      &                        &                        & 64.79       & 69.03       \\\cline{1-1}\cline{3-3}\cline{6-7}
        {\bf UNIT} \cite{LiuMY01}      &                                 & 23.46                      &                        &                        & 71.63       & 74.58       \\\cline{1-1}\cline{3-3}\cline{6-7}
        {\bf CycleGAN} \cite{ZhuJ01}   &                                 & 52.41                      &                        &                        & 71.53       & 85.83       \\\cline{1-1}\cline{3-3}\cline{6-7}
        {\bf MI{$^2$}GAN} (Ours)       &                                 & {\bf 72.86}                &                        &                        & {\bf83.49}  & {\bf 94.87} \\\cline{1-1}\cline{3-3}\cline{6-7}
        {\bf MI{$^2$}GAN w/o $D_{MI}$} &                                 & 65.96                      &                        &                        & 77.27       & 92.17       \\\hline
    \end{tabular}
\end{table}

\begin{figure}[!tb]
    \begin{center}
        \includegraphics[width=0.85\linewidth]{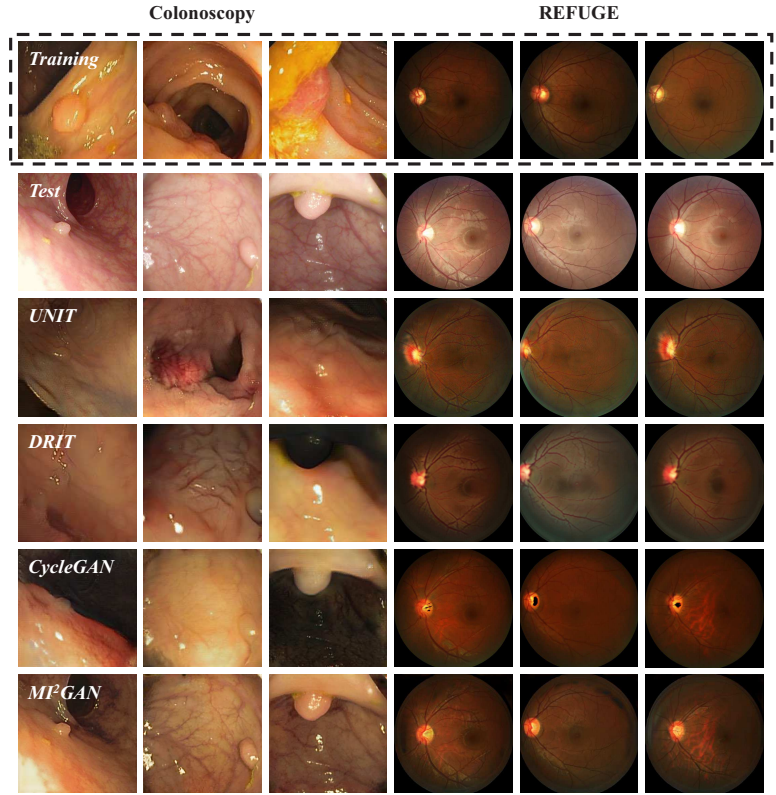}
    \end{center}
    \caption{Comparison of images produced by different I2I adaptation approaches.}
    \label{translation_res}
\end{figure}

\subsection{Comparison to State of the Art}
Different I2I domain adaptation approaches are applied to the colonoscopic and fundus image datasets, respectively, which translate the test images to the domain of the training set to narrow the gap between them and improve the model generalization. The adaptation results generated by different I2I domain adaptation approaches are presented in Fig.~\ref{translation_res}. The first row of Fig.~\ref{translation_res} shows the examplars from the training sets of colonoscopy and REFUGE datasets. Content distortions are observed in the adaptation results produced by most of the existing I2I translation approaches. In contrast, our MI{$^2$}GAN yields plausible adaptation results while excellently preserving the image-contents.

For quantitative analysis, we present the segmentation accuracy of deep learning networks with different adaptation approaches in Table~\ref{table_colon_simple}. To comprehensively assess the adaptation performance of our MI{$^2$}GAN, we adopt two widely-used deep learning networks, i.e., ResUNet-50 \cite{He01,Ronneberger01} and DeepLab-V3 \cite{deeplabv32017}, for the polyp segmentation, and OC/OD segmentation, respectively. As shown in Table~\ref{table_colon_simple}, due to the lack of capacity of image-content preservation, most of existing I2I domain adaptation approaches degrade the segmentation accuracy for both tasks, compared to the direct transfer. The DRIT \cite{Lee01} yields the highest degradation of DSC, $-40.87\%$, $-16.87\%$ and $-24.46\%$ for polyp, OC and OD, respectively. Conversely, the proposed MI{$^2$}GAN remarkably boosts the segmentation accuracy of polyp ($+8.53\%$), OC ($+1.83\%$), and OD ($+1.38\%$) to the direct transfer, which are closed to the accuracy on validation set.

\section{Conclusion}
In this paper, we proposed a novel GAN (namely MI$^2$GAN) to maintain image-contents during cross-domain I2I translation. Particularly, we disentangle the content features from domain information for both the source and translated images, and then maximize the mutual information between the disentangled content features to preserve the image-objects.

\section*{Acknowledge}
This work is supported by the Natural Science Foundation of China (No. 91959108 and 61702339), the Key Area Research and Development Program of Guangdong Province, China (No. 2018B010111001), National Key Research and Development Project (2018YFC2000702) and Science and Technology Program of Shenzhen, China (No. ZDSYS201802021814180).

\bibliographystyle{splncs04}

\end{document}